\documentclass[sigconf, nonacm]{acmart} 

\usepackage{booktabs}
\usepackage{tabularx}

\AtBeginDocument{%
  }

\begin{document}

\title{A Bounded Reclaim Actuator for PSI-Guided Compressed Memory: A Controlled Ablation}

\author{Abhiyan Dhakal}
\email{itsabhiyandhakal@gmail.com}
\orcid{0009-0003-3143-416X}
\affiliation{%
  \institution{Department of Computer Science and Engineering, Kathmandu University}
  \city{Dhulikhel}
  \country{Nepal}
}

\author{Sanjog Sigdel}
\correspondingauthor
\email{sanjog.sigdel@ku.edu.np}
\orcid{0009-0003-3143-416X}
\affiliation{%
  \institution{Department of Computer Science and Engineering, Kathmandu University}
  \city{Dhulikhel}
  \country{Nepal}
}

\renewcommand{\shortauthors}{Dhakal et al.}

\begin{abstract}
When the aggregate working set of active processes exceeds physical RAM capacity, the machine experiences memory pressure. Applications may therefore slow down before the kernel kills a process. Linux provides several ways to observe and respond: Pressure Stall Information (PSI) can detect memory-related task stalls, zram can provide compressed in-memory swap space, and cgroup v2 can request memory reclamation within a selected control group. These facilities are often discussed together even though they act at different points in the pressure path. This paper examines that distinction with a controlled systems study. We compare three setups: zram enabled from startup; zram enabled only after PSI indicates memory pressure; and zram enabled from startup with a one-time 96 MiB cgroup reclaim request. We first selected the request size in a 16-case pilot, then ran 180 confirmatory cases, 60 cases for each setup, on nine 1-vCPU Linux virtual machines with compute and SQLite workloads. Compared with static zram, the bounded reclaim configuration reduced compute p99 response time by 6\%, while the SQLite result was statistically indistinguishable. Delayed activation had higher median p99 latency than both alternatives. These results suggest that the benefit depends on the foreground workload and its memory-access path, rather than a general improvement across workloads.
\end{abstract}

\begin{CCSXML}
<ccs2012>
   <concept>
       <concept_id>10011007.10010940.10010941.10010949.10010950.10010951</concept_id>
       <concept_desc>Software and its engineering~Virtual memory</concept_desc>
       <concept_significance>500</concept_significance>
       </concept>
 </ccs2012>
\end{CCSXML}

\ccsdesc[500]{Software and its engineering~Virtual memory}

\keywords{
Memory Pressure, Pressure Stall Information (PSI), cgroup v2, zram, Proactive Memory Reclaim, Operating System}
  
\maketitle

\section{Introduction}
\label{sec:intro}

Memory pressure is a condition in the operating system where the demand for random-access memory (RAM) approaches or exceeds the available physical memory. This forces the system to reclaim memory or use slower storage backed mechanisms such as swapping. A foreground request can spend time waiting for a page to be reclaimed, written to swap, or read back, even while the machine is still alive and making progress. For a user, it is a slow interaction. Compressed memory is attractive because it can make a small amount of physical memory behave like a larger tier. However, the compression backend does not decide which pages should leave the active set or when that work should happen \cite{douglis1993compression,wilson1999compressed,tuduce2005adaptive,cervera1999swapcompression}.

In Linux, Pressure Stall Information (PSI) reports time stalled while tasks wait for memory or other resources \cite{linux-psi,weiner2018psi}. zram provides a compressed block device that the kernel can use as swap \cite{linux-zram}; zswap is a related compressed cache placed in front of a conventional swap device \cite{linux-zswap}. A control group is a Linux mechanism for placing processes in a hierarchy and applying resource limits and accounting to them. The cgroup v2 interface is the current unified version of that mechanism. Its \texttt{memory.reclaim} control file accepts a byte amount written by a user space program and asks the kernel to reclaim approximately that much memory from the selected cgroup \cite{linux-cgroup}. The kernel retains control over page selection.

This paper compares three configurations under the same memory-pressure workload to inspect whether a user space program can use a sustained PSI signal to request bounded reclaim and improve foreground responsiveness when compressed swap is already available. First, zram is available from the beginning of the run. Second, zram is enabled only after a sustained PSI signal. Third, zram is available from the beginning and the same PSI signal triggers one 96 MiB reclaim request against the background cgroup.

\section{Related Work}
\label{sec:related}

Compressed memory has been used to extend effective memory capacity and reduce access to slower storage \cite{douglis1993compression,wilson1999compressed,tuduce2005adaptive}. Swap-compression studies similarly showed that compression can improve application performance when physical memory is insufficient \cite{cervera1999swapcompression}. More recent systems optimize lower memory tiers directly: ZNSwap co-designs swap with zoned storage, while STYX offloads zswap compression to a SmartNIC \cite{bergman2022znswap,ji2023styx}. These systems change storage placement or compression execution rather than the timing of a userspace reclaim request.

Other systems protect interactive work by changing reclaim, page preparation, or writeback. Redline treats responsiveness as an explicit resource-management objective \cite{redline2008}. Acclaim studies reclaim and refault costs that affect Android applications, ASAP prepares pages before application switches, and SWAM revises mobile swap and out-of-memory policy \cite{liang2020acclaim,son2021asap,lim2023swam}. PMR separates page shrinking from storage-friendly writeback, while Archer uses page associations to adapt compression for mobile response time \cite{li2025pmr,li2025archer}. These systems make stronger choices about page identity or scheduling than the cgroup request evaluated here.

Feedback-driven systems connect memory-pressure observations to repeated control decisions. TMO uses its Senpai userspace controller to monitor PSI and adjust memory offloading over time \cite{weiner2022tmo}. Hermit applies feedback-directed asynchrony to remote memory, PageFlex exposes userspace paging policies through eBPF, and DAMON and Multi-Gen LRU provide richer information about page access or recency \cite{qiao2023hermit,yelam2025pageflex,linux-damon-reclaim,linux-mglru}. These systems demonstrate a broader range of feedback, paging, and page-selection policies than the intervention studied here.

\section{Methodology}
\label{sec:method}

\begin{figure*}[t]
\centering
\includegraphics[width=0.91\textwidth]{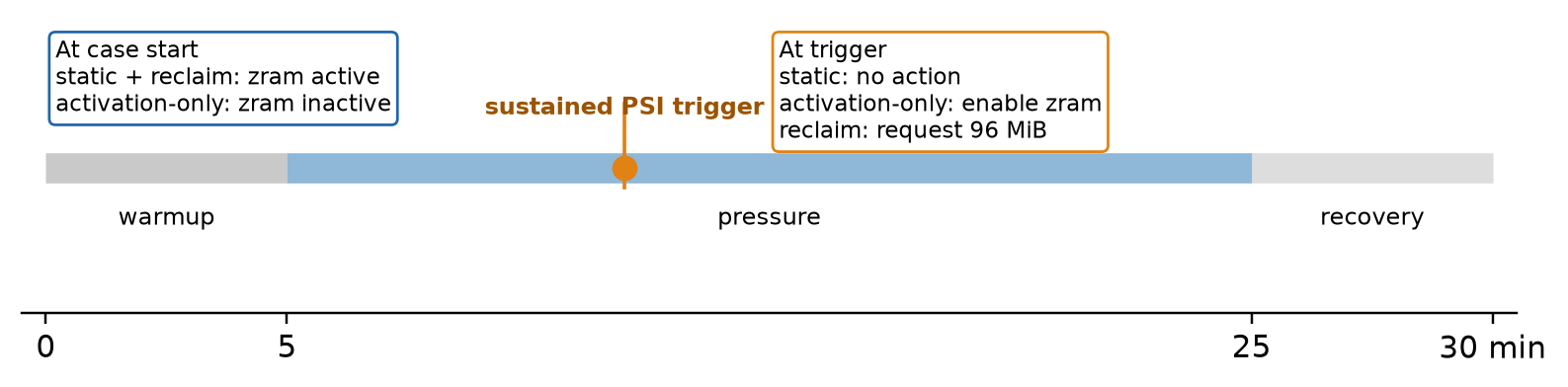}
% \caption{Shared case timeline and the two points at which the configurations differ. The trigger marker is schematic; its measured time varied by case.}
\caption{Shared case timeline with configuration-specific actions at a case-dependent trigger.}
\label{fig:timeline}
\end{figure*}

Each experimental case combined one configuration with the background pressure generator and both foreground workloads. Every case followed warmup, pressure, and recovery phases in the sequence shown in Figure~\ref{fig:timeline}.

\subsection{Experimental Environment}

The confirmatory experiment ran on qualified 1-vCPU, 1-GiB Linux VMs denoted \texttt{vm0} through \texttt{vm8}. Before scheduling cases, qualification checked root access, the kernel and experiment configuration, a writable \texttt{memory.reclaim} interface, zram, a swapfile, available disk space, clean starting swap, and the source commit used for the campaign.

Each case created sibling background and foreground cgroups. The background cgroup allocated and repeatedly touched memory to create pressure. The foreground cgroup ran the compute and SQLite workloads. The reclaim configuration targeted only the background cgroup, so the controller did not directly reclaim the foreground workloads' memory. Table~\ref{tab:config} summarizes the configuration held constant across the confirmatory cases.

\begin{table}[h]
\caption{Fixed configuration used in all confirmatory cases.}
\label{tab:config}
\centering\footnotesize
\begin{tabularx}{\columnwidth}{@{}lX@{}}
\toprule
Item & Value \\
\midrule
Case phases & 5-min warmup, 20-min pressure, 5-min recovery \\
Background memory & 384 MiB limit; 640 MiB target; 128 MiB active set \\
Compressed swap & LZO zram; 1-GiB device; 20\% memory limit \\
\bottomrule
\end{tabularx}
\end{table}

\subsection{Pressure and Foreground Workloads}

The pressure generator generated an allocation larger than the background cgroup's limit while repeatedly touching its active set. Earlier sensitivity tests showed that allocation rate changed the interval between PSI and material swap growth, while page contents changed zram occupancy and writeback \cite{dhakal2026dataset}. The confirmatory design therefore crossed two representative allocation rates with structured and random contents, producing the four regimes in Table~\ref{tab:regimes}. Table~\ref{tab:config} gives the settings held constant.

\begin{table*}[h]
\caption{The four pressure regimes used in the confirmatory study.}
\label{tab:regimes}
\centering\small
\begin{tabular}{@{}llllr@{}}
\toprule
Regime & Allocation step & Nominal rate & Page contents & Nominal allocation time \\
\midrule
Moderate/structured & 16 MiB every 200 ms & 80 MiB/s & Compressible structured bytes & 8 s \\
Moderate/random & 16 MiB every 200 ms & 80 MiB/s & Low-compressibility random bytes & 8 s \\
Gradual/structured & 16 MiB every 1 s & 16 MiB/s & Compressible structured bytes & 40 s \\
Gradual/random & 16 MiB every 1 s & 16 MiB/s & Low-compressibility random bytes & 40 s \\
\bottomrule
\end{tabular}
\end{table*}

The foreground cgroup ran two canaries, small periodic workloads used to observe latency during pressure. The compute canary maintained a hot memory region and performed memory touches and computation. The SQLite canary issued indexed point queries against an 80,000-row database with 1,536-byte payloads; SQLite ran as an embedded library inside the canary process rather than as a separate server. Both canaries ran every 100 ms and recorded the scheduled start, actual wake time, and completion time. Wake delay is the interval from scheduled start until execution begins, while service time is the interval from that point until completion. Each trace also recorded deadline misses, faults, and context switches. A deadline miss occurred when a request did not complete before its next 100 ms period.

\subsection{Configurations and PSI Trigger}

The static-zram configuration enabled the fixed backend before the case began and performed no controller action. The delayed-activation configuration began with that backend inactive and enabled it once after the trigger. The bounded-reclaim configuration enabled the backend before the case began and wrote 96 MiB once to the background cgroup's \texttt{memory.reclaim} file after the trigger.

Linux memory PSI distinguishes two stall states. The \texttt{some} state accumulates time during which at least one non-idle task in the monitored cgroup is stalled on memory, whereas \texttt{full} accumulates time during which all non-idle tasks are stalled simultaneously. Linux maintains a cumulative \texttt{total} counter for each state in microseconds \cite{linux-psi}.

The two triggered configurations used the same predicate. The controller read the background cgroup's PSI counters in one-second windows and required the \texttt{some total} counter to increase by at least 1 ms in each of three consecutive windows. It then performed the action assigned to the configuration; static zram had no trigger action.

\subsection{Pilot Dose Selection}

The pilot selected the reclaim amount before the confirmatory schedule was generated. It ran on \texttt{vm0} with a 2-minute warmup, 5-minute pressure phase, and 2-minute recovery. The four candidate requests, 16, 32, 64, and 96 MiB, were each tested in the four pressure regimes, giving 16 pilot cases.

A dose was eligible only if all four requests completed within one second, produced no OOM or interface error, achieved at least 80\% of the requested amount computed as \texttt{pgsteal} times the page size, and left valid backend and restoration telemetry. Table~\ref{tab:pilot} reports the worst duration and minimum achieved/requested ratio observed for each dose. All four doses passed, so the predefined rule selected the largest eligible request, 96 MiB.

\begin{table}[h]
\caption{Pilot results used to select the 96 MiB reclaim request.}
\label{tab:pilot}
\centering\small
\begin{tabular}{@{}lrrrr@{}}
\toprule
Dose & Max. duration & Min. achieved/requested & Errors & Decision \\
\midrule
16 MiB & 73.6 ms & 1.0010 & 0 & Eligible \\
32 MiB & 51.5 ms & 1.0000 & 0 & Eligible \\
64 MiB & 118.5 ms & 1.0003 & 0 & Eligible \\
96 MiB & 164.4 ms & 1.0009 & 0 & Selected \\
\bottomrule
\end{tabular}
\end{table}

\subsection{Confirmatory Schedule}

The confirmatory dataset contains 60 matched triplets, each containing one case per configuration under the same VM, pressure regime, and repeat identifier. These triplets produced 180 cases. Every case retained its original outcome. A case could be rerun only after an infrastructure failure, not after an OOM, failed reclaim, partial reclaim, or unfavorable latency result.

\subsection{Statistical Analysis}

The two primary outcomes were pressure-phase p99 response time for the compute and SQLite workloads. For each workload, bounded reclaim was compared with static zram and delayed activation, giving four primary comparisons. Using the matched triplets, the analysis computed the logarithm of each reclaim-to-comparator response-time ratio and resampled repeats within each VM, regime stratum. Strata received equal weight before the estimate was transformed back to a ratio; values below one favor bounded reclaim.

Each primary comparison reports a simultaneous 98.75\% bootstrap interval. This interval is an uncertainty range for the ratio, and the simultaneous level is chosen so that the four intervals together have a 95\% family-wise target rather than treating four separate tests as unrelated. The predefined rule required all four upper bounds to be below one and all four point estimates to show at least a 5\% reduction before claiming a general responsiveness benefit. An interval entirely above one indicated harm for that comparison. Absolute p95 and p99 differences, wake and service time, deadline misses, PSI, refaults, swap, zram, and reclaim execution were secondary measurements. An absolute difference is the latency in one configuration minus the latency in the other, in microseconds; it complements the scale-free ratio by showing the size of the change in the observed units.

% \subsection{Reproducibility}

% The complete pilot and confirmatory measurements, analysis exports, and reproduction instructions are available with the submission artifacts.

% The checksum-indexed Zenodo v2 release contains the complete pilot and confirmatory measurements \cite{dhakal2026datasetv2}. The accompanying analysis exports provide case summaries, primary ratios, mechanism measurements, and the data used to generate the figures and tables in this paper. Detailed archive verification and reproduction instructions remain with the dataset and repository rather than in the main research narrative.

\section{Evaluation}
\label{sec:evaluation}

\begin{figure*}[t]
\centering
\includegraphics[width=0.7\textwidth]{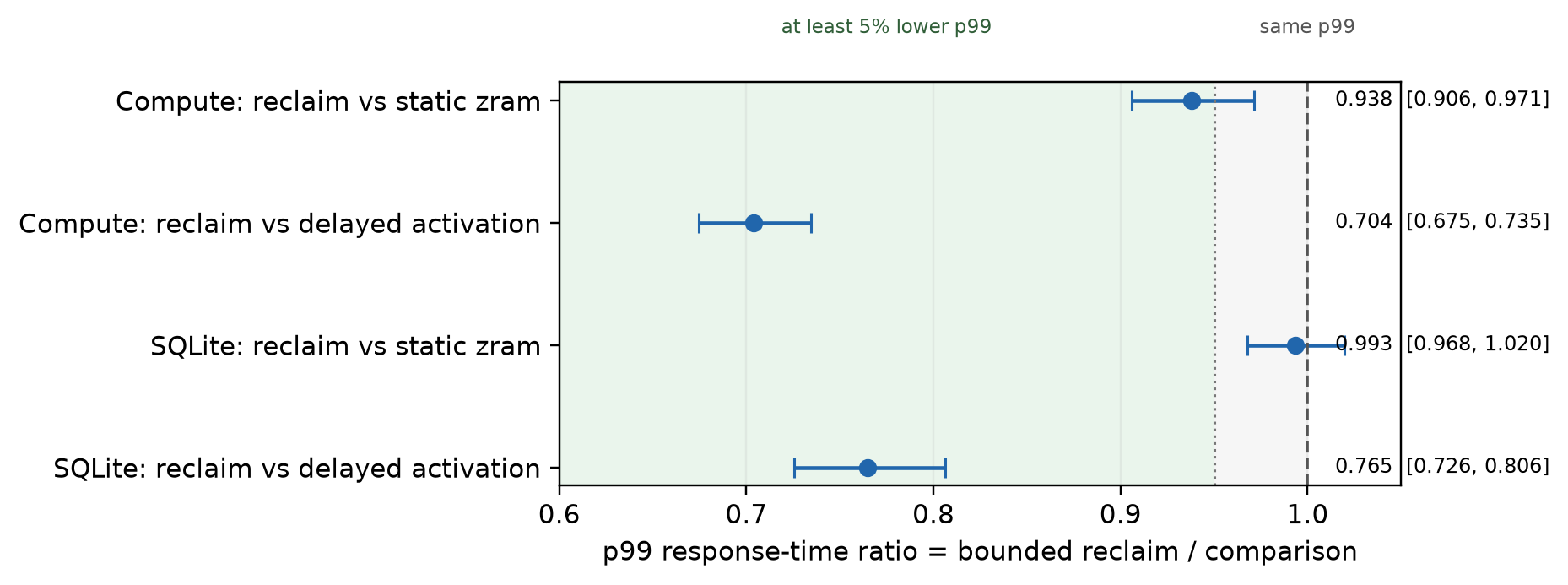}
%\caption{The four prespecified pressure-phase p99 comparisons. Each point is bounded reclaim divided by the named comparison; values below 1 mean lower p99 under reclaim. Horizontal lines are simultaneous 98.75\% paired bootstrap intervals. An interval crossing 1 indicates that the data do not show a clear difference.}
\caption{Prespecified pressure-phase p99 ratios for bounded reclaim versus the two comparison configurations.}
\label{fig:primary}
\end{figure*}

All scheduled confirmatory cases were available for analysis, and no primary comparison had a missing outcome. Figure~\ref{fig:primary} summarizes the four primary response-time ratios. All 60 bounded-reclaim requests completed successfully, with a median duration of 131.5 ms, p95 duration of 178.7 ms, and a minimum achieved/requested ratio of 1.0000.

% \begin{table*}[h]
% \caption{The 96 MiB request completed successfully in all 60 reclaim cases.}
% \label{tab:mechanism}
% \centering\small
% \begin{tabular}{@{}lrrrrrr@{}}
% \toprule
% Contents & Requests & \shortstack{Median duration\\(ms)} & \shortstack{P95 duration\\(ms)} & \shortstack{Median achieved/\\requested} & \shortstack{Min. achieved/\\requested} & Failures \\
% \midrule
% All reclaim cases & 60 & 131.5 & 178.7 & 1.0021 & 1.0000 & 0 \\
% Structured contents & 30 & 128.7 & 178.9 & 1.0019 & 1.0000 & 0 \\
% Random contents & 30 & 136.1 & 168.4 & 1.0025 & 1.0005 & 0 \\
% \bottomrule
% \end{tabular}
% \end{table*}

\subsection{Foreground Response Time}

Table~\ref{tab:primary} reports the four primary comparisons. For the compute workload, the reclaim-to-static ratio was 0.938 with interval [0.906, 0.971], corresponding to an estimated 6.2\% reduction in p99 response time. The reclaim-to-delayed-activation ratio was 0.704 [0.675, 0.735], also favoring reclaim.

For SQLite, reclaim improved p99 relative to delayed activation: the ratio was 0.765 [0.726, 0.806]. The comparison with static zram was 0.993 [0.968, 1.020]. Because this interval includes one, the data do not show a clear SQLite difference between those two configurations. The predefined all-four rule therefore reports no general responsiveness benefit and no detected harm.

\begin{table*}[h]
\caption{Primary p99 comparisons between bounded reclaim and the two alternatives.}
\label{tab:primary}
\centering\small
\begin{tabular}{@{}llrrrr@{}}
\toprule
Workload & Comparison & Ratio & Lower & Upper & Median absolute difference ($\mu$s) \\
\midrule
Compute & Reclaim/static & 0.938 & 0.906 & 0.971 & -12.5 \\
Compute & Reclaim/delayed activation & 0.704 & 0.675 & 0.735 & -384.5 \\
SQLite & Reclaim/static & 0.993 & 0.968 & 1.020 & +36.5 \\
SQLite & Reclaim/delayed activation & 0.765 & 0.726 & 0.806 & -380.0 \\
\bottomrule
\end{tabular}
\end{table*}

\subsection{Wake and Service Time}

Table~\ref{tab:decomp} separates the case-level wake and service measurements across the 60 cases in each configuration. Delayed activation's larger response times coincided with larger wake-delay medians. Bounded reclaim did not reduce every component relative to static zram: for example, its median compute and SQLite wake p99 values were slightly higher. The compute improvement therefore should not be described as a uniform speedup of every timing component.

Deadline misses were rare in all three configurations. This controlled workload can expose tail differences, but the low miss counts do not establish that any configuration would protect deadlines for a larger interactive application.

\begin{table*}[h]
\caption{Foreground latency and deadline misses under each configuration.}
\label{tab:decomp}
\centering\small
\resizebox{\textwidth}{!}{\begin{tabular}{@{}lrrrrrrr@{}}
\toprule
Configuration & Compute p95 ($\mu$s) & Compute p99 ($\mu$s) & Compute wake p99 ($\mu$s) & SQLite p95 ($\mu$s) & SQLite p99 ($\mu$s) & SQLite wake p99 ($\mu$s) & Deadline misses \\
\midrule
Static zram & 373.5 & 943.0 & 778.5 & 859.5 & 1574.0 & 794.5 & 3 \\
Delayed activation & 482.0 & 1337.5 & 1187.0 & 1081.0 & 2008.5 & 1247.0 & 4 \\
Bounded reclaim & 369.0 & 921.0 & 792.0 & 864.5 & 1631.5 & 813.5 & 3 \\
\bottomrule
\end{tabular}}
\end{table*}

\subsection{Pressure and Memory Behavior}

The mechanism measurements describe how each configuration changed the pressure episode. Median accumulated \texttt{some} stall time and \texttt{full} stall time were 1.342 s and 1.114 s with bounded reclaim, 1.515 s and 1.306 s with delayed activation, and 1.818 s and 1.730 s with static zram. Median refault counts were 32,786, 32,905, and 34,104, respectively. Median swap-out was also close for bounded reclaim and static zram (99,027 and 99,198 KiB), while delayed activation reached 103,028 KiB and showed the expected late growth in zram occupancy after its backend was enabled.

These counters explain system behavior but do not replace the foreground measurements. A PSI reduction shows that the pressure path changed, and a completed reclaim request shows that the kernel performed work, but neither proves that a particular foreground workload benefited. The SQLite/static comparison demonstrates this distinction: the mechanism executed successfully without producing a clear p99 improvement for that workload.

\subsection{Variation Across Pressure Regimes}

Table~\ref{tab:regime-results} reports descriptive case-level medians for each treatment within each pressure regime; each cell summarizes 15 cases. The values expose the workload dependence hidden by an overall median: bounded reclaim was not the lowest treatment in every cell. These unpaired medians describe the cells; the prespecified paired analysis in Table~\ref{tab:primary} determines the primary conclusions.
Delayed activation had the highest median p99 in every regime, while bounded reclaim alternated with static zram depending on workload and regime.

\begin{table}[h]
\caption{Foreground p99 across workloads and pressure regimes.}
\label{tab:regime-results}
\centering\small
\setlength{\tabcolsep}{4.4pt}
\begin{tabular}{@{}lrrrrrr@{}}
\toprule
& \multicolumn{3}{c}{Compute p99 ($\mu$s)} & \multicolumn{3}{c}{SQLite p99 ($\mu$s)} \\
\cmidrule(lr){2-4}\cmidrule(l){5-7}
Regime & Static & Act. & Reclaim & Static & Act. & Reclaim \\
\midrule
Moderate/structured & 1,044 & 1,440 & 1,014 & 1,727 & 2,184 & 1,697 \\
Moderate/random & 908 & 1,290 & 1,002 & 1,533 & 1,901 & 1,724 \\
Gradual/structured & 895 & 1,249 & 911 & 1,372 & 1,916 & 1,666 \\
Gradual/random & 927 & 1,339 & 775 & 1,495 & 2,014 & 1,485 \\
\bottomrule
\end{tabular}
\end{table}

\section{Limitations}
\label{sec:limitations}

The homogeneous 1-vCPU experimental population reduces configuration variation and supports paired comparisons, but it does not represent larger servers, desktop systems, mobile devices, or heterogeneous production fleets. The reported intervals describe this population rather than all Linux systems.

The background allocator was designed to produce repeatable pressure, and the two foreground workloads were designed to expose wake and service delays. They do not reproduce the full behavior of an interactive desktop, database server, or distributed application. The difference between the compute and SQLite results is consistent with workload-dependent memory access, but the experiment does not identify every cause of that difference.

The study evaluated only the pilot-selected request and issued it once. That fixed action may not be appropriate for other memory sizes or workloads. The experiment does not test adaptive doses, repeated feedback, or a controller that changes its action as pressure evolves.

Because victim selection remained under kernel control, the request and mechanism counters cannot show which pages created the later latency behavior. Establishing that causal path would require additional page-level tracing or an intervention that controls victim selection.

The three-configuration comparison does not establish performance against the complete policies reviewed in Section~\ref{sec:related}. Five repeat identifiers per regime also limit stratum-specific precision. Finally, the inconclusive SQLite/static interval is not evidence that the configurations are equivalent; it means that this study did not detect a difference under the predefined test.

\section{Conclusion}
\label{sec:conclusion}

This study compared static zram, delayed activation, and bounded reclaim using compute and SQLite foreground canaries.

Delayed activation had the highest descriptive median p99 for both workloads. Bounded reclaim reduced compute p99 relative to static zram and reduced both workloads' p99 relative to delayed activation. It did not produce a clear SQLite improvement over static zram, so the predefined decision rule did not establish a general responsiveness benefit.

The workload-dependent result motivates adaptive repeated control, broader workloads and machine sizes, page-protection or victim-selection mechanisms, and further study of proactive reclaim and writeback.

\bibliographystyle{ACM-Reference-Format}
\bibliography{references}

\end{document}